\documentclass[%
 reprint,
 superscriptaddress,
 amsmath,
 amssymb,
 aps
 ]{revtex4-2}

\usepackage{graphicx}
\usepackage{dcolumn}
\usepackage{bm}
\usepackage{textgreek}
\usepackage{braket}
\usepackage{dsfont}
\usepackage{array}
\usepackage{booktabs}
\usepackage{rotating}
\usepackage{gensymb}
\usepackage[colorinlistoftodos,prependcaption,textsize=tiny]{todonotes}
\usepackage{nicefrac}
\usepackage{mathtools}
\usepackage{hyperref}
\usepackage{bbm}

\begin{document}
\preprint{APS/123-QED}

\title{Composable free-space continuous-variable quantum key distribution using discrete modulation}

\author{Kevin Jaksch}
\email{kevin.jaksch@mpl.mpg.de}
\affiliation{Max Planck Institute for the Science of Light, Staudtstr. 2, 91058 Erlangen, Germany}
\affiliation{Department of Physics, Friedrich-Alexander-Universit\"at Erlangen-N\"urnberg, Staudtstr. 7, 91058 Erlangen, Germany}
\affiliation{SAOT, Graduate School in Advanced Optical Technologies, Paul-Gordan-Str. 6, 91052 Erlangen, Germany}

\author{Thomas Dirmeier}
\affiliation{Max Planck Institute for the Science of Light, Staudtstr. 2, 91058 Erlangen, Germany}
\affiliation{Department of Physics, Friedrich-Alexander-Universit\"at Erlangen-N\"urnberg, Staudtstr. 7, 91058 Erlangen, Germany}

\author{Yannick Weiser}
\affiliation{Max Planck Institute for the Science of Light, Staudtstr. 2, 91058 Erlangen, Germany}
\affiliation{Department of Physics, Friedrich-Alexander-Universit\"at Erlangen-N\"urnberg, Staudtstr. 7, 91058 Erlangen, Germany}

\author{Stefan Richter}
\affiliation{Max Planck Institute for the Science of Light, Staudtstr. 2, 91058 Erlangen, Germany}
\affiliation{Department of Physics, Friedrich-Alexander-Universit\"at Erlangen-N\"urnberg, Staudtstr. 7, 91058 Erlangen, Germany}

\author{\"Omer Bayraktar}
\affiliation{Max Planck Institute for the Science of Light, Staudtstr. 2, 91058 Erlangen, Germany}
\affiliation{Department of Physics, Friedrich-Alexander-Universit\"at Erlangen-N\"urnberg, Staudtstr. 7, 91058 Erlangen, Germany}

\author{Bastian Hacker}
\affiliation{Max Planck Institute for the Science of Light, Staudtstr. 2, 91058 Erlangen, Germany}
\affiliation{Department of Physics, Friedrich-Alexander-Universit\"at Erlangen-N\"urnberg, Staudtstr. 7, 91058 Erlangen, Germany}

\author{Conrad R\"oßler}
\affiliation{Max Planck Institute for the Science of Light, Staudtstr. 2, 91058 Erlangen, Germany}
\affiliation{Department of Physics, Friedrich-Alexander-Universit\"at Erlangen-N\"urnberg, Staudtstr. 7, 91058 Erlangen, Germany}

\author{Imran Khan}
\affiliation{Max Planck Institute for the Science of Light, Staudtstr. 2, 91058 Erlangen, Germany}
\affiliation{Department of Physics, Friedrich-Alexander-Universit\"at Erlangen-N\"urnberg, Staudtstr. 7, 91058 Erlangen, Germany}

\author{Stefan Petscharning}
\affiliation{AIT Austrian Institute of Technology, Center for Digital Safety\&Security, Giefinggasse 4, 1210 Vienna, Austria}

\author{Thomas Grafenauer}
\affiliation{AIT Austrian Institute of Technology, Center for Digital Safety\&Security, Giefinggasse 4, 1210 Vienna, Austria}

\author{Michael Hentschel}
\affiliation{AIT Austrian Institute of Technology, Center for Digital Safety\&Security, Giefinggasse 4, 1210 Vienna, Austria}

\author{Bernhard \"Omer}
\affiliation{AIT Austrian Institute of Technology, Center for Digital Safety\&Security, Giefinggasse 4, 1210 Vienna, Austria}

\author{Christoph Pacher}
\affiliation{AIT Austrian Institute of Technology, Center for Digital Safety\&Security, Giefinggasse 4, 1210 Vienna, Austria}
\affiliation{fragmentiX Storage Solutions GmbH, xista science park, Pl\"ocking 1, 3400 Klosterneuburg, Austria}

\author{Florian Kanitschar}
\affiliation{AIT Austrian Institute of Technology, Center for Digital Safety\&Security, Giefinggasse 4, 1210 Vienna, Austria}
\affiliation{Vienna Center for Quantum Science and Technology (VCQ), Atominstitut, Technische Universit\"at Wien, Stadionallee 2, 1020 Vienna, Austria}
\affiliation{Institute for Quantum Computing and Department of Physics and Astronomy, University of Waterloo, Waterloo, Ontario, Canada N2L 3G}

\author{Twesh Upadhyaya}
\affiliation{Institute for Quantum Computing and Department of Physics and Astronomy, University of Waterloo, Waterloo, Ontario, Canada N2L 3G}

\author{Jie Lin}
\affiliation{Institute for Quantum Computing and Department of Physics and Astronomy, University of Waterloo, Waterloo, Ontario, Canada N2L 3G}

\author{Norbert L\"utkenhaus}
\affiliation{Institute for Quantum Computing and Department of Physics and Astronomy, University of Waterloo, Waterloo, Ontario, Canada N2L 3G}

\author{Gerd Leuchs}
\affiliation{Max Planck Institute for the Science of Light, Staudtstr. 2, 91058 Erlangen, Germany}
\affiliation{Department of Physics, Friedrich-Alexander-Universit\"at Erlangen-N\"urnberg, Staudtstr. 7, 91058 Erlangen, Germany}

\author{Christoph Marquardt}
\email{christoph.marquardt@fau.de}
\affiliation{Department of Physics, Friedrich-Alexander-Universit\"at Erlangen-N\"urnberg, Staudtstr. 7, 91058 Erlangen, Germany}
\affiliation{Max Planck Institute for the Science of Light, Staudtstr. 2, 91058 Erlangen, Germany}

\date{\today}

\begin{abstract}
Continuous-variable (CV) quantum key distribution (QKD) allows for quantum secure communication with the benefit of being close to existing classical coherent communication. In recent years, CV QKD protocols using a discrete number of displaced coherent states have been studied intensively, as the modulation can be directly implemented with real devices with a finite digital resolution.
However, the experimental demonstrations until now only calculated key rates in the asymptotic regime. To be used in cryptographic applications, a QKD system has to generate keys with composable security in the finite-size regime.
In this paper, we present a CV QKD system using discrete modulation that is especially designed for urban atmospheric channels. For this, we use polarization encoding to cope with the turbulent but non-birefringent atmosphere. This will allow to expand CV QKD networks beyond the existing fiber backbone. In a first laboratory demonstration, we implemented a novel type of security proof allowing to calculate composable finite-size key rates against i.i.d. collective attacks without any Gaussian assumptions. We applied the full QKD protocol including a QRNG, error correction and privacy amplification to extract secret keys. In particular, we studied the impact of frame errors on the actual key generation.
\end{abstract}

\maketitle

\section{\label{intro} Introduction}
Continuous-variable (CV) quantum key distribution (QKD) can use squeezed or coherent states \cite{Ralph1999,GG02,Silberhorn2002} and homodyne detection to allow for quantum secure communication between two parties whilst being close to existing classical coherent communication \cite{Khan2015,Pirandola2020Review,Zhang2024}. 
For long, research focused on Gaussian modulated protocols, where coherent states are displaced according to a 2D Gaussian distribution. This allowed to exploit the optimality of Gaussian attacks and to simplify the security proof techniques \cite{Laudenbach2018}. 
This lead to tremendous advances in fiber based systems, demonstrating long-distance CV QKD \cite{Jouguet2013Nat,Zhang2020} even for true local oscillators (LO) \cite{Hajomer2024LLO}, and embedding the systems in composable frameworks against collective attacks for both practical block sizes \cite{Jain2022} and resistant against modulation imperfections \cite{Hajomer2022}.

However, the continuous Gaussian distribution remains a theoretical idealization, as experimental implementations are limited by the finite discretization of the used electronic devices. Thus, discrete-modulated (DM) systems with a finite number of displaced coherent states have been investigated \cite{Wittmann2010,Heim2014,Kleis2017}. Yet, one cannot assume the optimality of Gaussian attacks for discrete modulation. Without that, the security proofs had the challenge of working in the infinite dimensional Hilbert space populated by coherent states. 

In recent years, new proof techniques have been able to overcome this limitation. For example, for asymptotic security proofs \cite{Ghorai2019,Denys2021}, there have been several experimental demonstrations using quadrature amplitude modulation (QAM) with 16 to 256 displaced states \cite{Roumestan2024,Pan2022,Hajomer2024Chip}. Here, no error correction was implemented yet and the authors hence only gave estimates on the achievable asymptotic key rates assuming error correction with an efficiency of $\beta = 95\,\%$ with respect to the Shannon limit of the classical channel between Alice and Bob. Another experiment also demonstrated a four-state (QPSK) protocol with implemented low-density parity-check (LDPC) codes for error correction \cite{Wang2022}. They however did not study the impact of the error correction on the real extractable key, but rather used the measured efficiencies of the LDPC codes to calculate asymptotic key rates again. In general, it occurs that many experimental papers in CVQKD (e.g. \cite{Tian2023,Xu2023}) deal with error correction by assuming a certain efficiency $\beta$ and adding the potential frame error rate (FER) as a scaling factor to the asymptotic key rate formula $K = \left(1 - \mathrm{FER} \right) \cdot \left(\beta I_\mathrm{AB} - \chi_\mathrm{BE} \right)$. This leads to the conclusion that higher key rates are obtained when operating the system close to the threshold of the used error correction code (ECC), leading to a high efficiency whilst increasing the FER. This approach has already raised concerns in recent years \cite{Johnson2017}.

Another proof technique for asymptotic DMCV QKD using a numerical approach \cite{Coles2016,Winick2018} was introduced in~\cite{Lin2019}. This approach has been proven to be very versatile, being able to include a trusted detector model \cite{Lin2020,Upadhyaya2021} and even composability against i.i.d. collective attacks \cite{Kanitschar2023}. Yet, an experimental implementation of this kind of protocols is missing and will be presented in this paper. Very recently, new theoretical studies have made partial progress towards proving security under coherent attacks \cite{Bauml2024,pascualgarcia2024,primaatmaja2024}, another work generalized this method to the multi-user scenario \cite{Kanitschar2024}.

All of the mentioned work so far was done for fiber-based systems, which can take advantage of the existing infrastructure in metropolitan areas. Complementary, free-space systems will allow to expand QKD networks beyond this existing but fixed fiber backbone. Here, the feasibility for CV QKD for both satellite-to-ground \cite{Gunthner2017,Dequal2021,Sayat2024Sat} and terrestrial urban links \cite{Heim2014,Shen2020}, as well as the general treatment of fading channels \cite{Usenko2012,Ruppert2019,Pirandola2021} has been studied intensively in recent years. With a full atmospheric implementation still missing, several laboratory studies to investigate technical aspects like rate adaptive error correction \cite{Gumus2024} or the effect of turbulences on wavefront distortions \cite{Sayat2024AO} show the vast interest in free-space applications.

In this work, we merge this recent endeavours in free-space and DMCV QKD. We developed a mobile QKD system that employs discrete QPSK modulation in the Stokes parameters. Using polarization modulation and Stokes detection allows for quantum communication in turbulent but non-birefringent atmospheric channels \cite{Heim2014} and is equivalent to homodyne detection using a transmitted LO \cite{Haeseler2008}. In our system, we also implemented the full QKD pipeline including a quantum random number generator (QRNG), error correction and privacy amplification. In a laboratory demonstration, we applied the security statement of \cite{Kanitschar2023} to calculate composable finite-size key rates with a total security parameter of $\epsilon = 1\cdot 10^{-10}$ against i.i.d. collective attacks. We further applied the post processing pipeline to generate secret keys and studied the effect of frame errors during error correction. For this, we do not just add a scaling factor to the key rate but stay in the framework of the given security statement. This shows that the key length is vulnerable to even small increases in the FER.

\section{\label{results}Results}
\subsection{Polarization-encoded modulation}
As the system is designed for atmospheric quantum communication, we exploit the non-birefringent nature of the atmosphere and use the polarization degree of freedom in order to encode the quantum states \cite{Heim2014}. This is different from amplitude and phase encoding, which is used in fiber-based systems and is subject to fluctuations in atmospheric turbulences. As described in this chapter, those two modulation schemes can be seen as equivalent, given that the local oscillator (LO) is not prepared locally at the receiver but sent through the quantum channel (transmitted LO scenario).

We prepare the bright LO and the coherent signal states in the same spatial mode but in orthogonal circular (left- and right-handed, L and R) polarization modes \cite{Korolkova2002}
\begin{equation}
    \ket{\psi} = \ket{\alpha_\mathrm{LO}}_\mathrm{L}\ket{\alpha_n}_\mathrm{R} , 
\end{equation}

with a discrete QPSK modulation for the signal states $\alpha_n = \lvert \alpha_n \rvert \cdot \exp{\left[\nicefrac{in\pi}{2}\right]}$, $n \in \left\{ 0,1,2,3 \right\}$ and $\lvert \alpha_\mathrm{LO} \rvert \gg \lvert \alpha_n \rvert$.

The polarization can thus be described using Stokes operators \cite{Korolkova2002},
\begin{equation}
\begin{aligned}
\hat{S}_0 &= \hat{a}^\dagger_\mathrm{L}\hat{a}_\mathrm{L} + \hat{a}^\dagger_\mathrm{R}\hat{a}_\mathrm{R} , &
\hat{S}_1 &= \hat{a}^\dagger_\mathrm{L}\hat{a}_\mathrm{R} + \hat{a}^\dagger_\mathrm{R}\hat{a}_\mathrm{L} , \\
\hat{S}_2 &= i\left(\hat{a}^\dagger_\mathrm{R}\hat{a}_\mathrm{L} - \hat{a}^\dagger_\mathrm{L}\hat{a}_\mathrm{R} \right) , &
\hat{S}_3 &= \hat{a}^\dagger_\mathrm{L}\hat{a}_\mathrm{L} - \hat{a}^\dagger_\mathrm{R}\hat{a}_\mathrm{R} ,
\end{aligned}
\end{equation}
where we express the mode operators in the circular basis with $\hat{\alpha}_\mathrm{R/L} = \nicefrac{1}{\sqrt{2}}\left( \hat{a}_x \pm i \hat{a}_y\right)$. 
As discussed in \cite{Haeseler2008}, the two-mode picture of Stokes operators is also the correct way to describe homodyne detection to measure the quadrature operators $\hat{X}/\hat{P}$ in a phase-encoded system for a LO transmitted through the untrusted optical channel. For that, one has to interchange the two orthogonal polarization modes with the two separate spatial modes of the LO and the signal beam. For the given scenario of a bright L-polarized LO $\hat{a}_\mathrm{L} \rightarrow \lvert \alpha_\mathrm{L} \rvert \exp{\left(i\phi\right)}$, this means that the Stokes detection of $\hat{S}_1$ and $\hat{S}_2$ corresponds to a quadrature measurement of $\hat{X}_R $ and $\hat{P}_R$ in the signal's orthogonal, R-polarized mode \cite{Josse2004}:
\begin{equation}
\label{Eq: StokesQuadratures}
\begin{aligned}
    \hat{S}_1 &\approx  \lvert \alpha_\mathrm{L} \rvert   \left(\hat{a}^\dagger_\mathrm{R} + \hat{a}_\mathrm{R} \right)= \sqrt{2} \, \lvert \alpha_\mathrm{L} \rvert  \hat{X}_\mathrm{R} ,\\
    \hat{S}_2 &\approx i  \lvert \alpha_\mathrm{L} \rvert   \left(\hat{a}^\dagger_\mathrm{R} - \hat{a}_\mathrm{R} \right)= \sqrt{2} \, \lvert \alpha_\mathrm{L} \rvert \hat{P}_\mathrm{R} . \\
\end{aligned}
\end{equation}
The relative phase $\phi$ is hereby set to zero since both polarization modes have no relative phase shift. 
Please note that the used approximation is the standard one for homodyne detection in the bright LO limit \cite{Leonhardt1995}.
The corresponding uncertainty relation is given by
\begin{equation}
    \mathrm{Var}\left( \hat{S}_1 \right) \cdot \mathrm{Var}\left( \hat{S}_2 \right) \geq \lvert \langle \hat{S}_3 \rangle  \rvert^2 ,
\end{equation}
being equal for coherent polarization states and increased in the presence of polarization excess noise.

\subsection{\label{experiment} Experimental implementation}
\begin{figure*}
\centering
\includegraphics[width=1\linewidth]{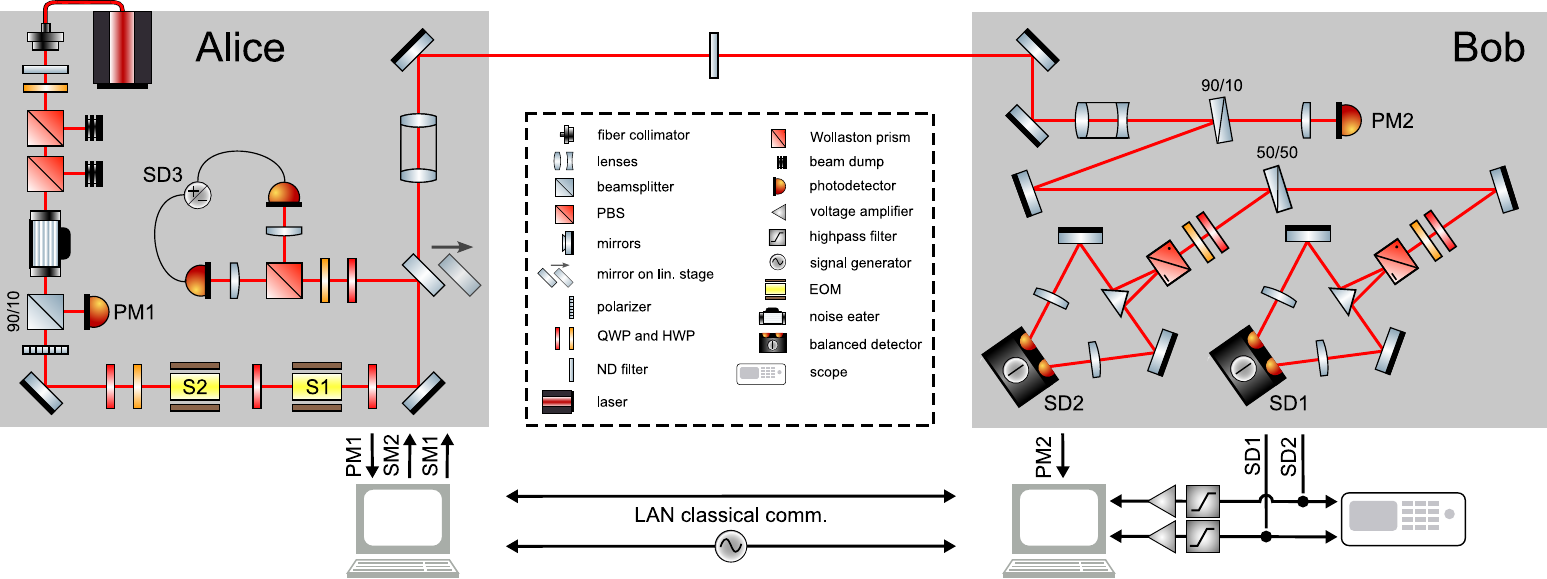}
\caption{\label{fig:setup} Experimental setup. At the sender, a circularly polarized bright LO with a discrete QPSK modulation in the Stokes ${S}_1$ and  ${S}_2$ parameters is prepared. At the receiver, the signal is measured by direct detection to monitor the transmission and two identical Stokes detection units, analogous to a heterodoyne measurement in optical phase space. The atmospheric channel is emulated by a 3\,dB ND filter.
Used abbreviations: PBS: polarizing beam splitter. QWP/HWP: quarter/half wave plate. ND filter: neutral density filter. EOM: electro-optical modulator. SM: Stokes modulation. SD: Stokes detection, PM: power monitoring}
\end{figure*}

The experimental setup consists of a breadboard-mounted and mobile sender (Alice) and receiver (Bob) module and is suitable for a flexible deployment in atmospheric links (see Fig. \ref{fig:setup}). The light of a continuous wave laser at 810\,nm is spatially mode-cleaned by a PM fiber and collimated into free-space. A stack of ND filters, a half-wave plate and polarizing beam splitters are used for coarse and fine power adjustment before the power is stabilized with a noise eater. A part of the beam is used to monitor the power on the sender side, the main fraction is sent through a combination of multiple wave plates and polarization modulators. By that combination, a circularly polarized strong LO with an additional and discrete modulation of four coherent states in the orthogonal polarization mode is prepared, equivalent to QPSK modulation in optical phase space. The signal at Alice can be measured with a first Stokes detection unit with two identical but separately amplified PIN photodiodes to quantify the laser's intensity noise by comparing the sum and difference signal.
For this first laboratory demonstration, we replaced an atmospheric channel with a 3\,dB ND filter. 

On Bob's side, a fraction of the beam is sent to a second monitoring detector to determine the transmission of the system. The remainder of the beam is split equally to two identical Stokes detection units to measure the Stokes $\hat{S}_1$ and  $\hat{S}_2$ operator. In contrast to Alice's side, Bob utilize balanced detectors that allow us to suppress correlated classical noise on the optical signal before amplification and further processing. The DC coupled output of the receiver is split and one part is monitored by a scope to check for imbalancing of the detectors. The rest of the signal is at first high-pass filtered at 130\,kHz to counteract low frequency drifts and then voltage amplified to match the dynamic range of the used analog-to-digital converter (ADC).
As further discussed in chapter \ref{tempmodes}, the RF spectra of the quantum signals is shifted away from the low frequency regime to avoid interference with the filter operation.
As a countermeasure for hacking attacks on the transmitted LO, we foresee to add a local vacuum calibration at random times and a Stokes $\hat{S}_3$ detection \cite{Haeseler2008} at Bob.

Clock synchronization between sender and receiver is achieved by a shared 10\,MHz signal from a function generator, that can be exchanged by two free-drifting Rubidium clocks at the sender and receiver, respectively, for future applications with separate locations. The classical communication between Alice and Bob is achieved via a LAN Ethernet connection.

\subsection{\label{modpattern}Signal preparation}
\begin{figure}
\centering
\includegraphics[width=1\linewidth]{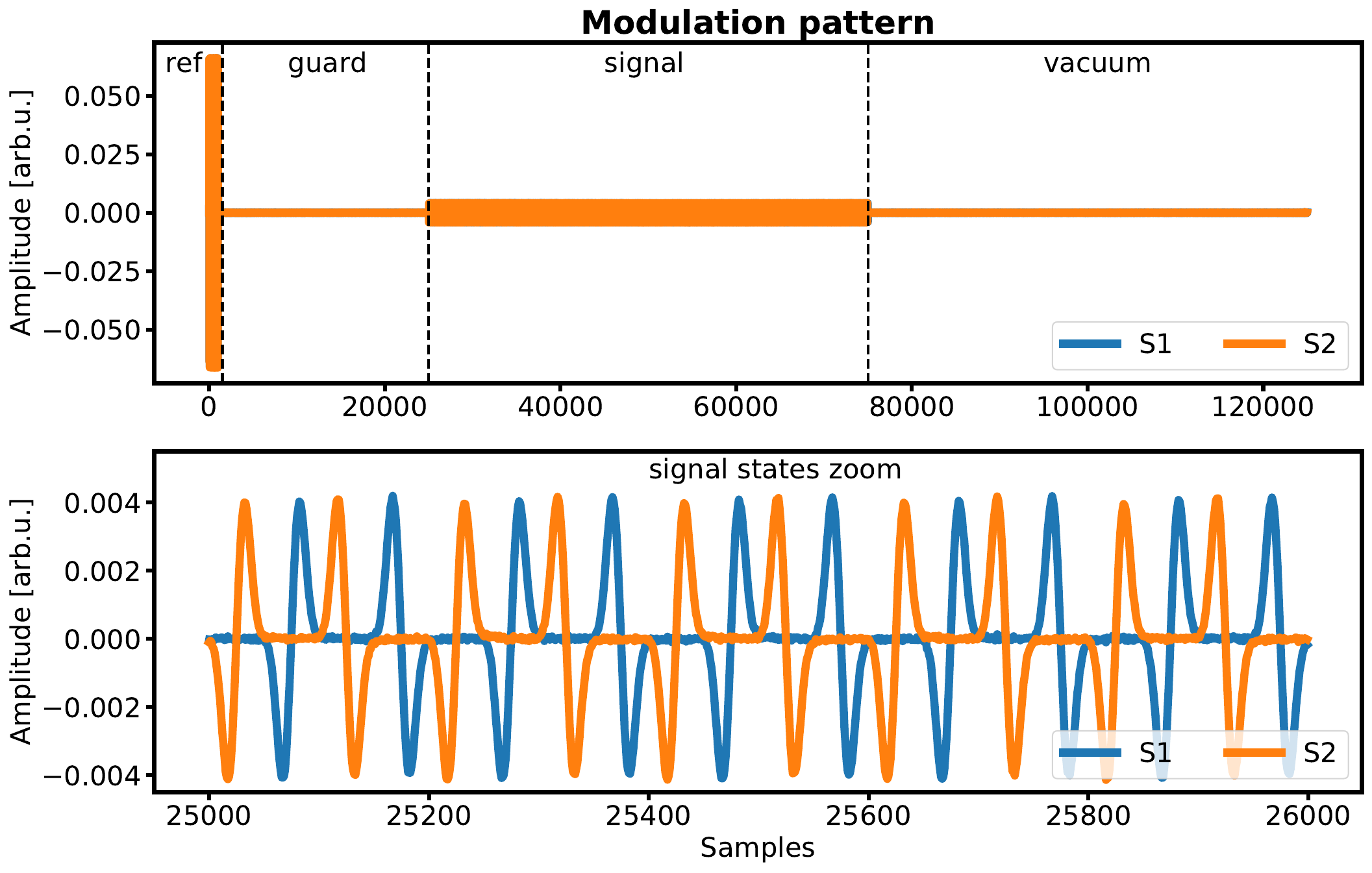}
\caption{\label{fig:ModPattern} Measured modulation pattern in the electrical domain as sent from the digital-to-analog converter with $f_{DAC} = 2$\,GHz. Top: Time-multiplexed frame with four slots for reference, guard, signal and vacuum states.
Bottom: Zoom into the signal states, encoded in Hermite-Gaussian pulses. For clarity, the plot is averaged over 240 frames with the same signal pattern.}
\end{figure}
The system is operated in a burst mode and its modulation pattern is designed to meet the requirements of future applications in atmospheric links. In a burst, Alice prepares 240 time-multiplexed frames each consisting of 2500 states modulated with 25\,MHz, as shown in Fig. \ref{fig:ModPattern}. Each state is then defined on a 40\,ns pulse window and temporally encoded in over-sampled Hermite-Gaussian pulses with different polarity and amplitudes (see chapter \ref{tempmodes}). The first 20 states are bright reference states with a fixed pattern used for triggering the ADC and frame synchronization. As the modulators may endure ringing or thermal effects after being driven by higher voltages, the next 480 states are neglected to avoid influence on the following 1000 signal states. For the signal, a random key string is provided by pre-saved random numbers from a fiber-integrated QRNG which is based on the balanced homodyne detection of the optical vacuum state \cite{Gabriel2010}. The entropy estimation was performed using a model considering a classical adversary \cite{Haw2015}. Two-universal hashing \cite{Carter1979} by means of the Toeplitz matrix method \cite{Mansour1990} was employed, in order to obtain random numbers with a statistical distance $\epsilon_\mathrm{Q}<2^{-100}$ to a uniform distribution. By using QPSK modulation, we can encode 2\,bits of information in each sent state (symbol). Finally, the modulation is turned off and the next 1000 states are used as a vacuum reference. Sending signal and vacuum states in the same frame copes with the fact that shot noise normalization is dependent on the power of the LO that will be sent through a fluctuating channel in atmospheric scenarios. As atmospheric turbulences are in the kHz regime \cite{Conan95}, we define a cut-off with a safety margin at 10\,kHz, so that the LO power in a 100\,\textmu s frame remains stable. This allows us to attribute a single transmission value to all the states of a frame, simplifying transmission monitoring and binning \cite{Usenko2012}.

\subsection{Signal processing}
After Bob's ADC is triggered, the measured trace is cross-correlated with the known pattern of the reference states in order to properly align the frames and pulse windows. As there are no phase fluctuations between the LO and the signal mode in the polarization based system, no additional phase correction step like in phase-encoded systems is necessary.
The individual quadrature values are then recovered from the over-sampled trace by calculating the weighted average over each pulse window and saved for further offline processing. Here, the weights are defined by the outgoing optical Hermite-Gaussian mode at the sender, which was characterized before executing the QKD protocol. 
The same procedure is performed for the signal, vacuum and detector states in order to have them defined according to the same temporal mode. The detector noise is measured in advance by blocking the laser beam at the receiver in the framework of a trusted detector assumption.

In following DSP steps, Alice uses further QRNG random numbers and announces 25\,\% of the sent key string. Bob normalizes his private and public quadrature string with respect to the measured vacuum variance (shot noise normalization). Bob further equalizes the trusted detector noise variances of both detectors by digitally adding white Gaussian noise to the acquired signal trace of the better detector with the lower detector noise. Adding such noise is a conservative step, making the performance worse while simplifying the DSP routine.
After having calculated the secret key rates, the private key strings are fed into the AIT-QPS pipeline for error correction and secret key extraction, as described in chapter \ref{ECC}.
For this demonstration, the quantum state exchange and the AIT-QPS pipeline were executed on separate computers using the LAN connection between Alice and Bob to emulate distant locations. Further offline processing steps and the key rate calculation were not yet performed locally.

\subsection{\label{tempmodes}Temporal mode}
\begin{figure}
\centering
\includegraphics[width=1\linewidth]{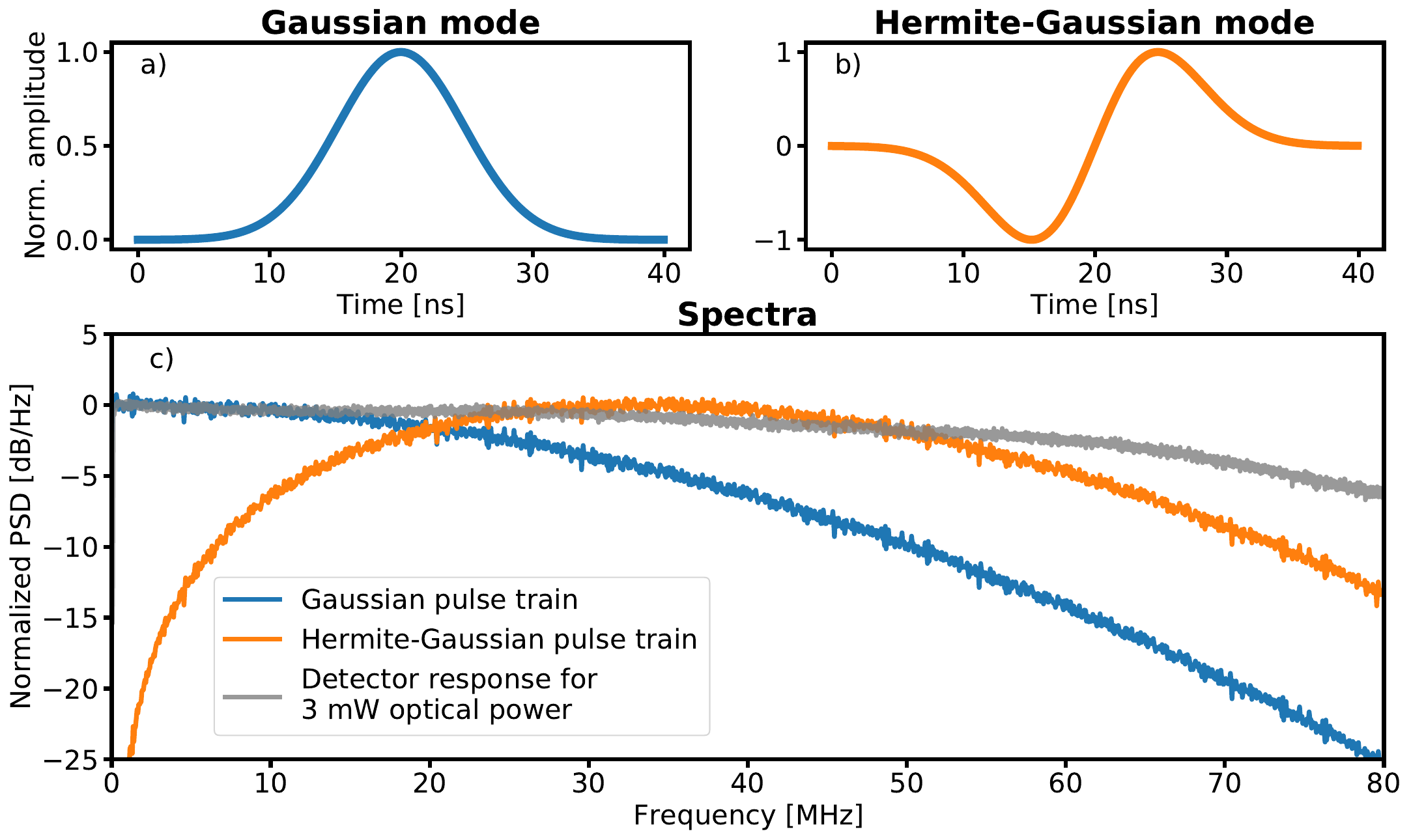}
\caption{\label{fig:tempmodes} Temporal modes in the time and spectral domain. a) Gaussian mode. b) First-order Hermite-Gaussian mode. c) RF spectra of a QPSK signal trace encoded in either Gaussian or Hermite-Gaussian pulses, showing the low frequency suppression of Hermite-Gaussian modes. The spectral response of the used homodyne detectors is flat and has a higher bandwidth than the QKD signal trace, allowing for proper mode conversion from the optical to the electrical domain. For the detector response, the electronic noise spectrum is subtracted.}
\end{figure}

As more and more advanced techniques from classical digital signal processing are applied to CV QKD systems in order to enhance their performance, the consequences of such algorithms on the underlying quantum optical definitions have to be taken into account. Especially, filter operations acting on the spectral properties of the signal are also changing the temporal mode and with that the definitions of the sent and measured quantum states. As outlined in the following, this could lead to a mode-mismatch between the sender and the receiver, which results in hidden additional detection losses, rendering estimated key rates insecure.

In this work, the quantum states are temporally encoded in modified Hermite polynomials of first order (Hermite-Gaussian pulses, see \cite{Micheal2002}) defined by
\begin{equation}
	\Gamma(t) = \frac{\sqrt{\mathrm{e}}\cdot t}{\sigma} \cdot \exp\left[-\frac{1}{2} \left( \frac{t}{\sigma}\right)^2\right] ,
\end{equation}
with $\sigma \approx 5$\,ns so that there is no interference between the different pulse windows. 
Similar to systems using frequency up-shifted RRC filtering \cite{Chin2018}, the RF spectra of such an encoded signal is moved away from the low frequency regime (see Fig. \ref{fig:tempmodes}).
The additional loss introduced by omitting the residual energy of the Hermite-Gaussian modes in the passband of the high-pass filters below 130\,kHz is in the order of $10^{-8}$ and currently neglected.
Furthermore, the pulse spectrum acts as an additional intrinsic filter against low frequency noise, reducing the preparation noise of the system by one order of magnitude with respect to a standard Gaussian pulse.

By defining the pulse shape, we also define the temporal mode $\Gamma(t)$ of the quantum states with the corresponding broadband-mode operators $\hat{A}^\dagger_\Gamma = \int \mathrm{d}t \ \Gamma(t) \hat{a}^\dagger(t)$ \cite{Brecht2015}. They also fulfill the bosonic commutation relation $\left[ \hat{A}^\dagger_\Gamma, \hat{A}_\Gamma \right] = 1$, allowing us to use the same framework as known from the standard single-mode treatment by replacing the single-mode ladder operators $\left\{\hat{a}^\dagger_{\omega}, \hat{a}_{\omega}\right\}$ with their broadband counterpart $\left\{\hat{A}^\dagger_\Gamma, \hat{A}_\Gamma\right\}$. However, one has to ensure that the measured quadrature operator $\hat{X}_\Lambda$ and its temporal mode $\Lambda(t)$ is defined according to the sent temporal mode, otherwise the mode mismatch between both modes leads to a decreased interferometric visibility and increased receiver losses.

For time-resolved homodyne detection, the measured mode is dependent on the temporal mode of the LO, the ADC oversampling, the bandwidth of the detectors, and the used linear DSP algorithm. We are using a continuous-wave LO and dense ADC oversampling with $f_{ADC} = 1.25$\,GHz. By taking the weighted average over the over-sampled pulse window and ensuring that the detector bandwidth is larger than the signal bandwidth, we can ensure a high mode-matching degree between both temporal modes \cite{Chen23}.

\subsection{\label{secproof}Security proof}

\begin{figure}
\centering
\includegraphics[width=1\linewidth]{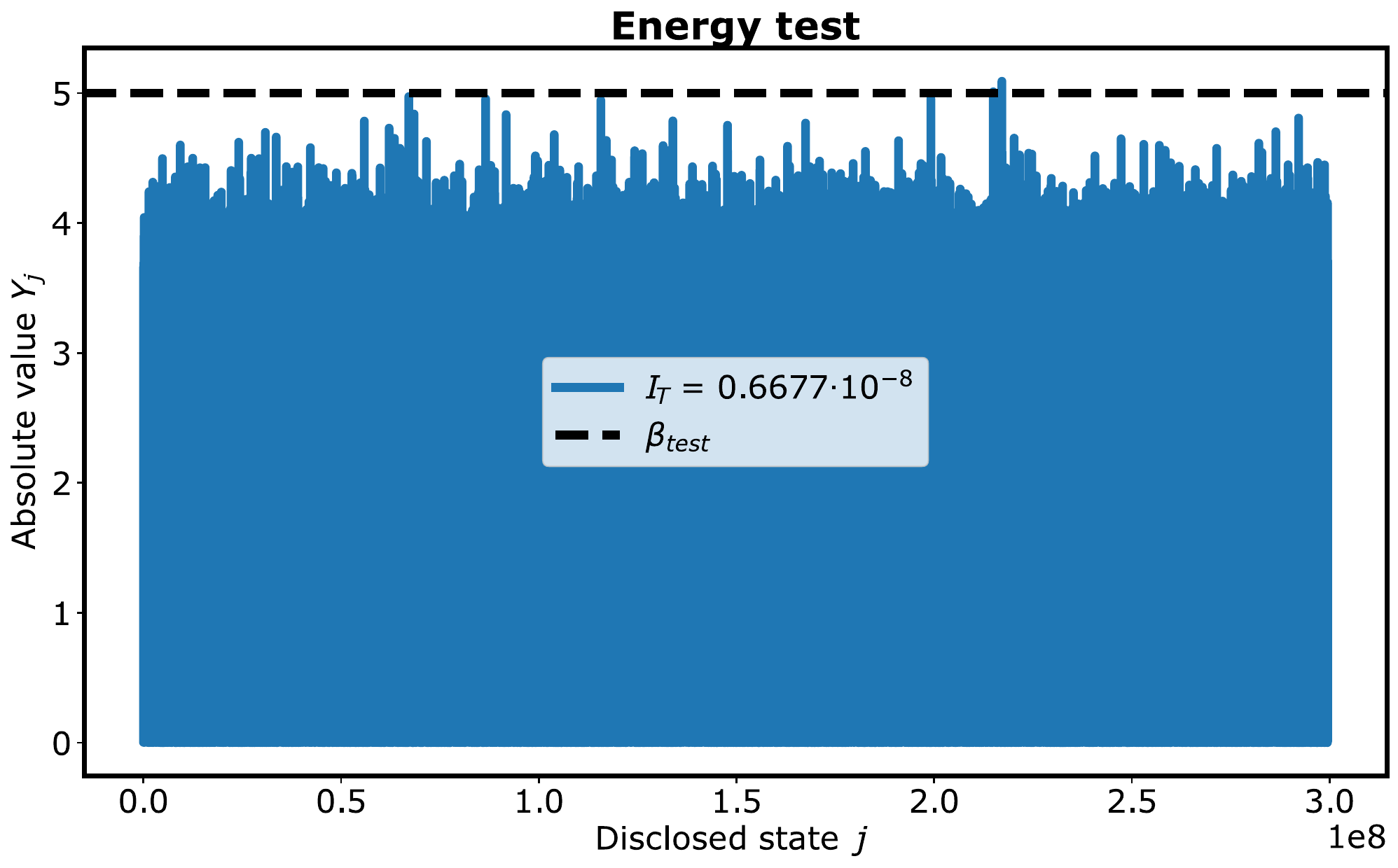}
\caption{\label{fig:ET} Energy test for Run 3, assuring that only a small number of outliers $l_T^\mathrm{meas}$ has an amplitude greater than $\beta_\mathrm{test} = 5$. With the chosen settings $n_c = 20$, $w=\mathcal{O}\left(10^{-7}\right)$, the energy test is passed if $I_T = \nicefrac{l_T^\mathrm{meas}}{k_T} < 1\cdot 10^{-8}$, allowing for maximal 2 outliers on the chosen disclosed block size $k_T$.}
\end{figure}

We discuss the underlying security argument along with the involved protocol steps and give application details for the security proof method of Ref. \cite{Kanitschar2023} used in this work. After Alice has prepared and distributed $N$ rounds of quantum states and Bob has performed his measurements, they proceed with statistical testing. In more detail, this means Alice and Bob disclose a random subset of size $k_T$ of their measurement outcomes publicly and use this information to perform an energy test and an acceptance test (see Refs. \cite{Kanitschar2023, Kanitschar_Thesis_2022} for details). The purpose of the energy test is the following: Coherent states, as prepared in the present protocol, consist of a linear combination of all photon number states. So, in particular, there is no maximal photon number that would allow us to represent them in a finite-dimensional space, making the underlying key rate finding problem amenable to numerical convex optimization algorithms. Therefore, we perform an energy test that allows us, based on experimental observations, to determine the weight of our quantum states outside a finite-dimensional cutoff space of dimension $n_c+1$. This weight then can be used to relate the actual infinite-dimensional optimization problem to a finite-dimensional cutoff formulation of the same problem, at the cost of introducing small weight-dependent correction \cite{Upadhyaya2021}. Practically, the test requires defining beforehand a cutoff dimension $n_c$, a testing parameter $\beta_{\text{test}}$, and a target weight $w$, as well as a number of allowed outliers $l_T$ (see Fig. \ref{fig:ET}). 
The target weight $w \in \left[0,1 \right]$ quantifies 'how much` of Bob's quantum state maximally may lie outside the chosen finite-dimensional cutoff space, $w\geq \text{Tr} \left[ \rho (\mathbbm{1}-\Pi)\right]$, where $\Pi$ denotes the projector on the finite-dimensional cutoff space of dimension $n_c+1$. Then, it counts the number of test rounds whose absolute value of the amplitude $Y_j$ exceeds $\beta_{\text{test}}$, $\l_T^{meas} = \#\{ Y_j: Y_j \geq \beta_{\text{test}} \} $. If this number is larger than the number of allowed outliers $l_T$, the test fails, except with probability $\epsilon_{\text{ET}}$, and the protocol aborts. Otherwise, we proceed with the acceptance test. During the acceptance testing procedure, we compare the observed averages of certain observables with a pre-defined set of accepted statistics. Based on the expected behavior of the QKD system, the acceptance set $\mathcal{S}^{\text{AT}}$ corresponding to the set of accepted statistics, contains all density operators that could have produced the expected statistics, except with probability $\epsilon_{\text{AT}}$. Thus, practically, the sole task during the acceptance test is comparing if the observed averages lie within the set of accepted statistics, and aborting otherwise. We want to highlight that the type of security argument used in this work does not rely on any assumptions about the quantum channel connecting Alice and Bob. In contrast, the security argument is based solely on the concept of testing based on Bob's measurement results without requiring any symmetry assumptions and rigorously handles infinite-dimensional quantum systems inherent to CV protocols, leaving merely the restriction to i.i.d. collective attacks.

\begin{figure}
    \centering
    \includegraphics[width=0.7\linewidth]{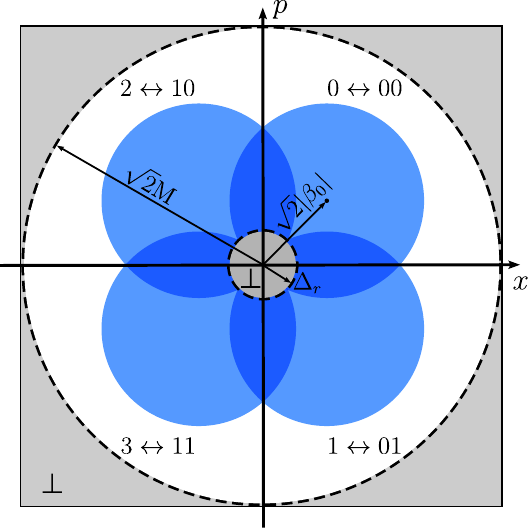}
    \caption{Sketch of the key map in phase space. The blue circles illustrate the uncertainty area of the received states with an amplitude $\beta_n = \sqrt{T\eta} \alpha_n$. Each measurement outcome is either discarded ($\perp$) or mapped to a two-bit value, defined by the projection on the axes. This maps the QPSK pattern to two binary symmetric channels on which error correction is performed. The security proof includes the possibility of radial postselection to enhance the performance. For this experiment, we chose $\Delta_r = 0$ and no postselection. Furthermore, detection events outside a bounded detection range $M =5$ (see Ref. \cite{Kanitschar2023}) are discarded.}
    \label{fig:Keymap}
\end{figure}

On the protocol level, we proceed with a (reverse reconciliation) key map, where Bob maps his measurement outcomes (which are complex numbers in phase space) to the set $\{0, 1, 2 ,3, \perp\}$, where $\perp$ stands for symbols discarded by postselection \cite{Silberhorn2002, Kanitschar2021}. Intuitively, this procedure is illustrated in Fig. \ref{fig:Keymap}. This is followed by error-correction, where Alice and Bob use the classical channel to reconcile their raw keys and error-verification, as well as privacy amplification, which helps them to decouple their key from Eve's information. Those steps are described in more detail in Section \ref{AITpipeline} and introduce the security parameters $\epsilon_{\text{EC}}$ and $\epsilon_{\text{PA}}$.

From a security proof perspective, we now have all the ingredients ready to formulate the security statement \cite{Kanitschar2023}. For $\epsilon_{\mathrm{ET}}, \epsilon_{\mathrm{AT}}, \bar{\epsilon}, \epsilon_{\mathrm{EC}}, \epsilon_{\mathrm{PA}} > 0$, the implemented QKD protocol is $\epsilon_{\mathrm{EC}} + \max\left\{\frac{1}{2}\epsilon_{\mathrm{PA}}+\bar{\epsilon}, \epsilon_{\mathrm{ET}}+\epsilon_{\mathrm{AT}} \right\}$-secure against i.i.d. collective attacks, given that, in case the protocol does not abort, the secure key length is chosen to satisfy
\begin{multline}
\label{secstatement}
    \frac{l}{N} \leq \frac{n}{N} \left[ \min_{\rho \in S^{E\&A}} H ( X \vert E )_\rho - \delta(\Bar{\epsilon}) - \Delta(w)  - 2\delta_{\mathrm{leak}} \right]\\ 
    - \frac{2}{N}\log_2\left( \frac{1}{\epsilon_{\mathrm{PA}}} \right)
\end{multline}
with $\delta_{\mathrm{leak}} = \nicefrac{\mathrm{leak}_{\mathrm{EC}}}{2n}$, where $\mathrm{leak}_{\mathrm{EC}}$ is the total number of disclosed bits, $\Delta(w) := \sqrt{w} \log_2(|Z|) + (1+\sqrt{w}) h\left( \frac{\sqrt{w}}{1+\sqrt{w}} \right)$, $\delta(\epsilon) := 2 \log_2\left( \mathrm{rank}(\rho_X)+3 \right) \sqrt{\frac{\log_2\left(2/\epsilon \right)}{n}}$ and $\mathcal{S}^{\mathrm{E\&A}}$ is the set of states that pass both the energy test and the acceptance test. By $|Z|$, we denote the dimension of Bob's classical system, which in this work is equal to $4$, and by $\rho_X$, we denote Alice's reduced density matrix after protocol execution.

While the error-correction leakage and the privacy amplification correction term are quantities determined by the respective software modules used, the first three terms in the key rate formula are obtained by the security argument. Since the weight was already fixed during the energy test and $\delta(\bar{\epsilon})$ is fixed upon choosing $\bar{\epsilon}$ (which is a ´virtual' parameter we could in principle optimize over), it remains to solve the finite-dimensional optimization problem $\min_{\rho \in S^{E\&A}} H ( X \vert E )_\rho$. The optimization over the set $\mathcal{S}^{\text{E\&A}}$ can be expressed as a non-linear semi-definite program, constrained by experimental observations along with further constraints ensuring we are optimizing over quantum states compatible with the protocol (see Supplementary for details). In particular, we chose the observables the displaced photon number $\hat{n}_{\beta_i}$ and the displaced squared photon number $\hat{n}_{\beta_i}^2$ for $i\in\{0,1,2,3\}$ (see Supplementary for details), which can be derived directly from Bob's heterodyne measurement results and are linked to the experimentally well-known quantities channel loss $\eta$ and excess noise $\xi$. The intuition behind this choice is the following; 
In case of no channel noise and that the displacement was chosen in accordance with the channel loss, the expectation for those observables is zero, while its deviation from zero is closely linked to the excess noise.
The analogous `displaced photon number basis' requires the least amount of basis states to describe the underlying quantum state up to fixed weight $w$, which makes it the natural choice.

This convex optimization problem can be solved via a method introduced in Refs. \cite{Coles2016, Winick2018} and applied to DM CV-QKD first in \cite{Lin2019}. It involves a two-step process, wherein the first step, a linearized version of the problem, is solved iteratively using the Frank-Wolfe algorithm \cite{Frank_Wolfe1956}. As it is not guaranteed to reach the exact minimum, the result of step 1 is merely an upper bound. However, it can serve as a starting point for step 2, where SDP duality theory and another linearization are combined to turn this upper bound into a reliable lower bound. A relaxation takes the effect of numerical constraint violations on the key rate into account. We used CVX \cite{cvx1, cvx2} to model the occurring SDPs as well as the MOSEK solver (Version 10.0.34) \cite{mosek} to solve them, while coding in \textsc{Matlab}\textsuperscript{\textregistered}, version 2022a. 
Additionally, we applied the techniques from \cite{Lin2020}, properly modeling and considering realistic, trusted detectors within the frame of this security proof technique. 
In particular, we want to highlight that the chosen security argument works directly with the discrete modulated signals without requiring any Gaussian assumption. The security proof further considers the exact constellation of every single symbol without relying on pattern averages or similar.

Please note that the system is using a two-mode Stokes measurement, whilst the security proof is based on a single-mode quadrature measurement. The security proof itself does not rely on any assumptions about the quantum channel. However, we are using the working assumption that the transmitted LO acting as the phase reference is unchanged, i.e., without an eavesdropper acting on it (honest implementation). This allows us to apply the single-mode security proof to our two-mode measurement. To formally overcome this mismatch, an additional Stokes  $\hat{S}_3$ detection is planned  \cite{Haeseler2008}. In addition, one needs to expand the security proof formally to the two-mode Stokes picture.

\subsection{\label{AITpipeline} AIT-QPS Pipeline}

The post-processing pipeline is a software consisting of several modules linked in series, where the key blocks are passed along, undergoing all required steps to distill a secure final key from the incoming raw key. An instance of the pipeline is launched at either site (Alice and Bob), running the same modules, only configured to act as the respective peer. The individual steps are described in the following.

Most modules require communication with their peer-module over a classical channel to exchange control information or key-related data. This channel is assumed to be error-free and completely public (e.g. an Ethernet connection). To prevent a man-in-the-middle attack, it needs to be authenticated. To this end, each key block is equipped with an authentication context, which gets passed along with the key throughout the whole process. In particular, all messages passed between peer-modules are hashed into the authentication context. For information-theoretical security, a universal hash function is used, the size of which is agreed upon beforehand. After processing of the key is concluded, a pre-shared secret is hashed onto the context, forming the final authentication tag. Now Alice and Bob compare their tags publicly to ensure authenticity. When the pipeline is running and the pre-shared secret is used up, it gets replenished by QKD keys produced by the pipeline itself. For this experiment, we restarted the pipeline for each run, always authenticating with the same initial pre-shared key.

At first, Bob’s measured continuous data is mapped to a two-bit value, defined by the projection on the axes of the optical phase space (see Fig. \ref{fig:Keymap}).
The following error correction is based on LDPC codes, operated in the binary-symmetric channel (BSC) model in a reverse reconciliation mode. An LDPC code is defined by the so-called Tanner-graph which is programmatically described by a sparse binary matrix. The multiplication of the input key with this matrix forms the syndrome (the encoded message). In reverse reconciliation, we arbitrarily define Bob’s key as correct, thus Bob encodes his key and sends the syndrome to Alice. The decoder algorithm at Alice tries to correct her key by means of belief-propagation in several iterations, ultimately matching her key to the syndrome. Each LDPC code is characterized by its code rate $r = \nicefrac{\left( L_\textrm{in} – L_\textrm{syn} \right) }{L_\textrm{in}}$, its  bit error rate (BER) threshold up to which at least 50\% of the keys can be corrected successfully and its input block length $L_\textrm{in}$. $L_\textrm{syn}$ is the syndrome length. In particular, we have used three codes with rates and thresholds r/th: 0.06125/0.3492 (ECC\#0), 0.07/0.3382 (ECC\#1), 0.08/0.3267 (ECC\#2).
The block length was 102400 bits for all codes, which is achieved by splitting up the complete input key block in multiple sub-blocks. The maximum number of iterations, after which the decoder declares a key as failed, was set to 400.

In rare cases a collision may occur, where different keys result in the same syndrome. In this case, the LDPC decoder will erratically declare a key as successfully corrected. To further reduce the probability of such cases, an additional step of confirmation is performed. To this end, a hash of pre-defined size is computed by both peers and compared. The composable finite-size security model requires, that all key material is passed on to privacy amplification, so failed blocks (both due to error correction as well as confirmation) are not discarded but openly exchanged and marked as such.

The process of privacy amplification entails the shortening of the key in order to render useless any information that may have leaked to an eavesdropper, both in the optical key exchange as well as during post-processing. It should be emphasized that, as required by the security model, prior to privacy amplification all key blocks are concatenated to form the initially sized input block. The size of the secure key is determined by the analysis following the security proof, in particular taking into account the amount of disclosed information during error correction and confirmation. For successfully corrected sub-blocks this is the syndrome size and the confirmation hash size, while for failed sub-blocks it is the total contained information. The actual process of key length reduction is implemented as a family of universal hash functions, one member of which is selected randomly. We use a Toeplitz matrix algorithm as hashing function, which is implemented by a fast polynomial multiplication.

\subsection{\label{keyrates} Key rate analysis}

\begin{figure}
\centering
\includegraphics[width=1\linewidth]{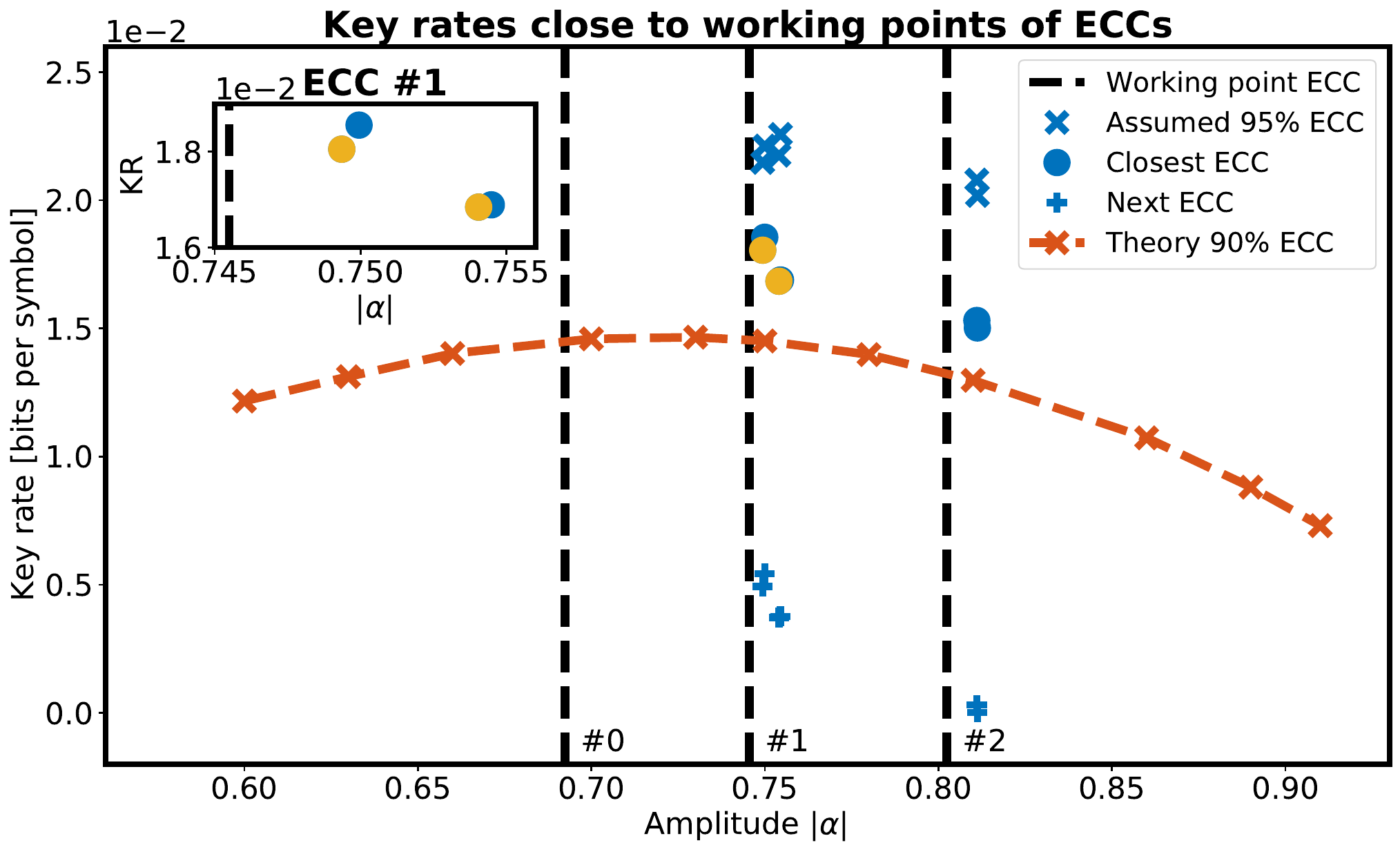}
\caption{\label{fig:keyrates} Composable finite-size key rates for either assuming a 95\,\% efficient error correction code or when applying different implemented LDPC codes. In order to find the optimal amplitude for the system, we precalculated key rates for conservative system parameters: $N = 1\cdot 10^9$, $r_{test} = 0.25$, $T = 0.49$, $\eta = 0.72$, $\nu_{el}  = 0.14$, $\xi_A = 0.3\,\%$, $\beta = 0.9$. The highlighted measurements are further analyzed in Fig. \ref{fig:ECC}.}
\end{figure}

We used the developed mobile sender and receiver QKD module in a first laboratory demonstration with an emulated 3\,dB loss channel and performed six measurements with a block size of $N = 1.20\cdot 10^9$ and a testing ratio of $r_{test} = 0.25$. Doing so, we acquired enough public states for statistically bounding the observable for the aforementioned finite-size security proof. Over all runs and symbols, the average experimental parameters of the system were given by the transmission $T = 0.494$, receiver efficiency $\eta = 0.720$, detector noise $\nu_{el} = 0.135\, \mathrm{SNU}$ and mean displaced (squared) photon number $\langle \hat{n}_{\beta} \rangle = 6.70\cdot 10^{-4}$ and $\langle \hat{n}^2_{\beta} \rangle = 6.77\cdot 10^{-3}$.  The measured $\langle \hat{n}_{\beta} \rangle$ is equivalent to an excess noise of $\xi_A = 2.71\cdot 10^{-3}\, \mathrm{SNU}$, defined at the channel input.
A detailed list of the measured parameters for all runs and symbols as used for the numerical SDP problem can be found in the Supplementary.

Being a first laboratory demonstration, we are characterizing the behavior of the QKD modules over a lossy channel without an eavesdropper acting on it. Doing so, we use the observed statistics of the six runs in this so-called honest implementation to both define the non-unique acceptance sets of the security proof and also to run the QKD protocol with them. In a practical implementation, one would have to test the measured observables of the untrusted channel against the beforehand defined acceptance sets. By optimizing the size of the acceptance set (by adjusting the t-vector in Theorem 3, \cite{Kanitschar2023}), we can optimize the protocol to have both a high key rate and a low abort-protocol probability for a passive Eve (protocol completeness). The computational time of the numerical SDP problem is thus shifted to the system’s characterization and acceptance set definition beforehand, leaving us with a fast abort/non-abort decision during the protocol run. The key rates could be further improved by using acceptance testing for variable-length security proofs, as recently proposed for discrete-variable QKD protocols with a framework open to expansion to the CV regime \cite{Tupkary2024}.

We operated the system at two different working points close to both the optimal amplitude maximizing the key rate and close to the implemented LDPC codes rendering efficient error correction and key generation feasible (see Fig. \ref{fig:keyrates}). For each of the six runs, the key rate was bounded with the security statement of Eq. \ref{secstatement} with the $\epsilon$-parameters set to $\epsilon_\mathrm{ET}=\nicefrac{1}{10}\cdot10^{-10}$, $\epsilon_\mathrm{AT}=\nicefrac{7}{10}\cdot10^{-10}$, $\bar{\epsilon} =\nicefrac{7}{10}\cdot10^{-10}$, $\epsilon_\mathrm{EC}=\nicefrac{2}{10}\cdot10^{-10}$ and $\epsilon_\mathrm{PA}=\nicefrac{1}{10}\cdot10^{-10}$, leading to a total security parameter of $\epsilon = \epsilon_{\mathrm{EC}} + \max\left\{\frac{1}{2}\epsilon_{\mathrm{PA}}+\bar{\epsilon}, \epsilon_{\mathrm{ET}}+\epsilon_{\mathrm{AT}} \right\}=  1\cdot 10^{-10}$. Furthermore, the key rate was calculated with three different error correction models. At first and as commonly used in the literature, the error correction costs $\delta_\mathrm{leak}$ were calculated by assuming a 95\,\% efficient ECC with respect to the BSC capacity the experimental bit error rate (BER). For the other two key rates, we don't assume the efficiency of a potential ECC. Instead, we subtracted the leakage of one of the implemented LDPC codes. For the second key rate, we used the LDPC code with its threshold BER closest to the measured BER (e.g. ECC\#1 for $\vert \alpha \vert \approx 0.75$). For the third key rate, we used the next, less optimal implemented code (e.g. ECC\#0 for the same amplitude).

The highest key rates were achieved when assuming a 95\,\% ECC, which is also the common figure of merit to compare the performance to other QKD systems without implemented error correction. Doing so, we achieved composable finite-size key rates up to $2.26\cdot 10^{-2}$ bits per symbol against collective i.i.d. attacks. Practically oriented, one has to study the key rates established with the true error correction costs. Since the implemented LDPC codes have a fixed code rate $ r = 1 - \delta_\mathrm{leak}$ with a fixed leakage $\delta_\mathrm{leak}$ per bit, their efficiency $\beta = \nicefrac{r}{c}$ reduces when increasing the amplitude and hence the channel capacity $c$ of the system. Consequently, the closest LDPC codes operating with an efficiency of approximately 89\,\% (see Table \ref{tab:keys}) achieve higher nominal key rates compared to the less optimal codes with a lower efficiency of approximately 78\,\%.
However, we have to operate the system close to the threshold BER of the LDPC codes in order to achieve high efficiencies. Being so close to its threshold, the error correction may fail to correct some of its blocks, leading to a non-zero FER.
As discussed in the next chapter, sacrificing the nominal key rate for the benefit of reducing the FER can optimize the actual extractable secret key after error correction.

\subsection{\label{ECC} Error correction and key extraction}
To study the impact of the implemented error correction and privacy amplification modules, we rewrite the key rate equation of the security statement in its leftover hashing lemma form, describing the extractable secret key length in bits:

\begin{multline}
\label{leftoverhasing}
    l \leq n \left[ \min_{\rho \in S^{E\&A}} H ( X \vert E )_\rho - \delta(\Bar{\epsilon}) - \Delta(\omega) \right]  - \mathrm{leak}_\mathrm{EC}\\ 
    - 2 \log_2\left( \frac{1}{\epsilon_{\mathrm{PA}}} \right)\,.
\end{multline}

Here, the conditional von Neumann entropy $H ( X \vert E )$ and both the correction terms for dimension reduction $\Delta(\omega)$ and the asymptotic equipartition property $\delta(\Bar{\epsilon})$ are still retrieved solving the security proof's SDP problem. The total error correction costs $\mathrm{leak}_\mathrm{EC}$ are now given by the number of disclosed bits in the LDPC module and the length of the confirmation hash $t=128$ bits. This leads to a correctness of the key of  $\epsilon_\mathrm{cor} = \epsilon_\mathrm{EC,imp} = \nicefrac{m}{t}\cdot 2^{-t} < 10^{-32}$, where m is the number of bits before error correction minus the hereby disclosed bits \cite{Krzic2023}. The secureness of the key is guaranteed by subtracting $ 2 \log_2\left( \nicefrac{1}{\epsilon_{\mathrm{PA}}} \right) = 100 $ bits from the secret key length in the privacy amplification module, leading to $\epsilon_\mathrm{PA,imp} = 2^{-50} $ and $\epsilon_\mathrm{sec} = \mathrm{max} \left\{ \nicefrac{1}{2} \,\epsilon_\mathrm{PA,imp} + \bar{\epsilon}, \epsilon_\mathrm{ET} + \epsilon_\mathrm{AT} \right\} = \nicefrac{4}{5} \cdot 10^{-10} $. For the whole implemented QKD protocol, including the QRNG with $\epsilon_\mathrm{Q} = 2^{-100}$,  we hence achieve a total security parameter of $\epsilon_\mathrm{imp} = \epsilon_\mathrm{Q} + \epsilon_\mathrm{sec} + \epsilon_\mathrm{cor} < 1\cdot 10^{-10}$. The results are listed in Table \ref{tab:keys}, where we achieve to generate composable secret keys with up to 1.6\,MB length for a finite block size of $N = 1.2\cdot 10^{9}$.

\begin{table}
\centering
\begin{tabular}{*7c}
\toprule
{} &  \multicolumn{3}{c}{Closest ECC} & \multicolumn{3}{c}{Next ECC}\\
\midrule
$\lvert \alpha \rvert $ & FER [\%] & $\beta$ [\%] & Key [MB] & FER [\%] & $\beta$ [\%] & Key [MB] \\
0.7494 & 16.00 & 89.48 & 0.19 & 0.00 & 78.29 & 0.74 \\
0.7499 & 14.37 & 89.42 & 0.52 & 0.00 & 78.24 & 0.81 \\
0.7540 & 5.89 & 88.23 & 1.60 & 0.00 & 77.20 & 0.55 \\
0.7545 & 6.18 & 88.77 & 1.56 & 0.00 & 77.67 & 0.56 \\
0.8111 & 5.07 & 88.13 & 1.38 & 0.00 & 77.12 & 0.04 \\
0.8112 & 9.79 & 88.29 & 0.49 & 0.01 & 77.25 & 0.00 \\
\bottomrule
\end{tabular}
\caption{Extracted secret keys and the influence of error correction for all six runs. Each run was either corrected with the ECC with a threshold BER closest to the experimental data or with the next, non-optimal ECC with a higher threshold BER. The longest key of 1.6 \,MB was extracted for moderating between FER and efficiency $\beta$. In particular, the key length is highly susceptible to frame errors.}
\label{tab:keys}
\end{table}

As stated in the introduction, it seems common in the literature to estimate the impact of frame errors by just adding a scaling factor $(1-\mathrm{FER})$ to the key rate equation. This would allow for optimizing key rates by operating the system close to the threshold of the error correction, increasing its efficiency $\beta$ on the cost of a high FER. However, one has to be careful on how frame errors are treated in the implemented QKD protocol and how that affects the different terms in the used security statement. In our case, in order to have a correct key shared between Alice and Bob, the blocks that have failed error correction have to be fully disclosed between them. This does not reduce the key length in the leftover hashing lemma (Eq. \ref{leftoverhasing}) with a scaling factor, but increases the number of disclosed bits $\mathrm{leak}_\mathrm{EC}$ that have to be subtracted from the secret key. In contrast to a pure scaling factor, this subtraction can quickly lead to zero key lengths when increasing the FER and hence disclosing too many blocks.

\begin{figure}
    \centering
    \includegraphics[width=1\linewidth]{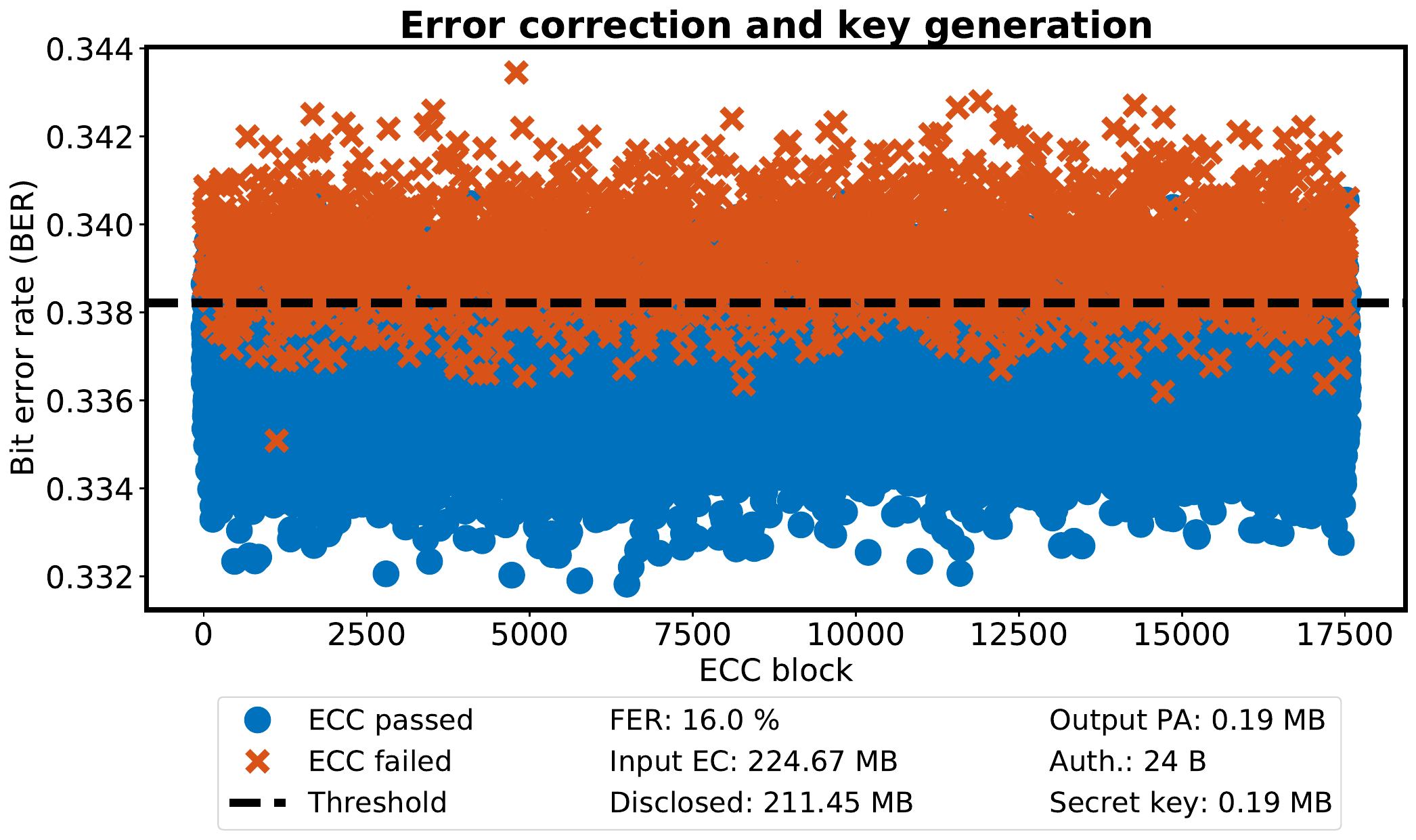}
    \includegraphics[width=1\linewidth]{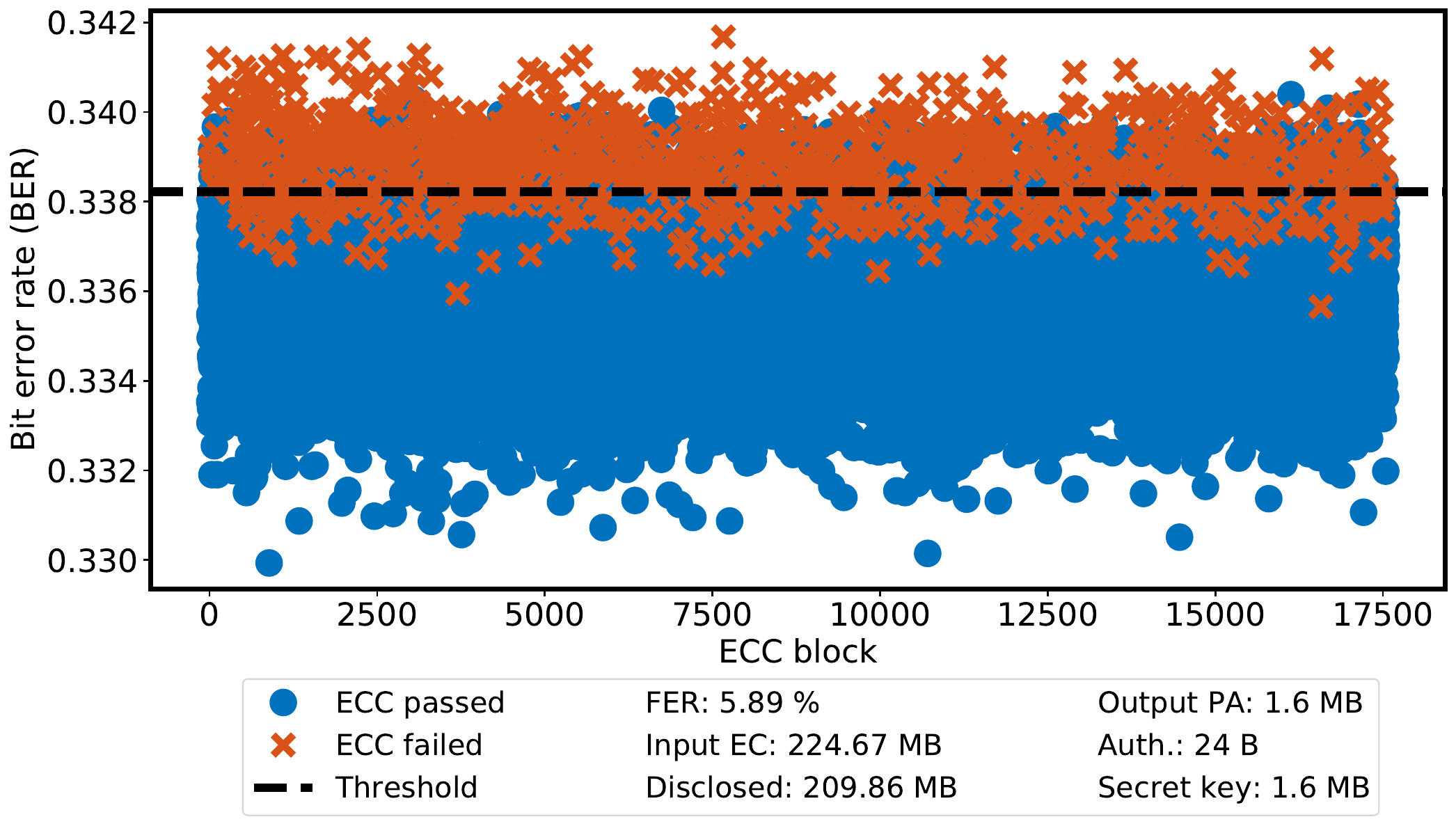}
    \caption{Detailed analysis of the postprocessing pipeline for run 1 (top) and run 3 (bottom) when applying the closest LDPC code ECC\#0. 
    Each block that cannot be corrected adds its whole length of 104000\,bits to the disclosed bits, reducing the extracted secret key. For each run, the AIT-QPS pipeline stores 2x96\,bits for the next authentication round.}
    \label{fig:ECC}
\end{figure}

As shown in Table \ref{tab:keys} for the first two runs, operating the system with LDPC codes with a high efficiency and FER provides less key than using a (at first glance non-optimal) LDPC code that only has an efficiency of 78\,\%, but can correct all blocks. For run 3 and 4, we slightly increased the amplitude to moderate between a lower FER with sufficiently high efficiency, leading to the highest extracted key lengths. We also operated the system at a different working point (run 5 and 6), where we also achieved high key lengths for moderate FERs. Please note the higher FER for the last run reducing the key length. This is due to a small drift in the modulators preparing the states, leading to a similar mean amplitude, but a deviation in the X/P bit streams after applying the key map projection. Like that, the P bit stream has a BER closer to the threshold BER, leading to more frame errors in correcting P (see Table 1 in the Supplementary). The effect of the frame errors is also illustrated in Fig. \ref{fig:ECC}.

\section{Discussion}
We present a CV QKD system using polarization encoding and discrete QPSK modulation designed for future applications in urban atmospheric channels. In a first proof-of-principle experiment in the laboratory, we implemented the protocol of a novel type of security proof that allows for composable finite-size key rates against i.i.d. collective attacks for discrete modulation without requiring any Gaussian assumptions.
We achieved composable finite-size key rates upon acceptance with up to  $2.26\cdot 10^{-2}$ bits per symbol. With the implemented QKD pipeline, including the QRNG, error correction and privacy amplification, we were able to generate secret keys up to 1.6\,MB for $N = 1.2\cdot 10^{9}$ sent states with a total security parameter of $\epsilon_\mathrm{imp} < 1\cdot 10^{-10}$.
We call the keys composable upon acceptance to highlight the fact that we used the honest implementation in the laboratory to both define the non-unique acceptance sets and run the QKD protocol with the statistics of the same measurement run. For a next practical implementation over an untrusted channel, we foresee testing the measured statistics against beforehand defined acceptance sets.

Having a transmitted LO is an active design choice for future atmospheric links, as LO and signal share the same spatial mode and will experience the same wavefront distortions. They are then auto-compensated when optically interfering during Stokes detection \cite{Heim2014}. Using a transmitted LO also allows for a low noise operation without the typical challenge of residual phase noise in systems with a true LO \cite{Hajomer2024LLO}.
However, to prevent hacking attacks on the transmitted LO \cite{Jouguet2013LO,Fan2023}, we foresee upgrading the power measurement to a Stokes $\hat{S}_3$ measurement as described in \cite{Haeseler2008}. Furthermore, we plan to add a second laser at Bob and switch to local vacuum measurements at random times, verifying the transmitted vacuum reference.

In the current system, the communication of the AIT-QPS pipeline is already authenticated using polynomial universal hashing with a pre-shared key \cite{Pacher2012}.  In order to authenticate the whole classical communication of the protocol, one has to use parts of the generated key for the authentication in the next run $n$ \cite{MullerQuade2009}. The new security parameter is then given by $\epsilon < n \left( \epsilon_\mathrm{imp} + \epsilon_\mathrm{auth} \right)$ with $\epsilon_\mathrm{auth} = \nicefrac{c}{a}\cdot 2^{-a}$ and a hash-size of $a =96$\, bits, linearly increasing with each run \cite{Krzic2023}. The length of the total classical communication (in bits) to be authenticated is denoted with $c$.  With $c \approx 450$\,MB in the AIT-QPS pipeline and $c \approx 30$\,MB during the optical quantum state exchange phase, including the latter in the authenticated communication will merely affect $\epsilon_\mathrm{auth} < 10^{-21}$.

We also studied the effect of frame errors on the extracted key length when not just using a scaling factor in front of the key rate, but when sticking to the security statement of the protocol. In the context of asymptotic key rates, our approach is similar to include the frame errors with $K = \left(1 - \mathrm{FER} \right) \beta I_\mathrm{AB} - \chi_\mathrm{BE}$, being seen as a lower bound when properly taking them into account \cite{Johnson2017}.
This treatment is indeed pessimistic as disclosed blocks currently do not add 0 bits to the key length, but actually add a negative contribution due to the fixed conditional entropy.
Another treatment could be to not disclose frame errors but to discard them completely. To do this properly, one has to include the postselection of frame errors in the postprocessing map of the protocol (see \cite{Lin2019}), similar to the already included phase space postselection. Discarded blocks would then not appear in the error correction costs, whilst being included in a modified smooth min-entropy.
A similar modification was done in the framework of composable security for Gaussian modulated protocols \cite{Pirandola2021,Jain2022}. 
Furthermore, using rate adaptive codes with a higher efficiency and lower FER will increase the performance of the system \cite{Zhou2021,Cil2024}. We also plan to study the effect of postselection, as a reduced data stream in the post processing pipeline will relax hardware requirements.
\\

\section*{Data availability}
The data that support the findings of this study are available from the corresponding authors upon reasonable request.

\section*{Acknowledgments}
We thank Lothar Meier and Oliver Bittel for their help in designing and building the used Stokes detectors.

This research was conducted within the scope of the project QuNET, funded by the German Federal Ministry of Education and Research (BMBF) in the context of the federal government’s research framework in IT-security ”Digital. Secure. Sovereign”.
F.K., T.U., J.L., and N.L. acknowledge the support from NSERC under the Discovery Grants Program (Grant No. 341495).

\section*{Author contributions}
K.J., T.D., and Y.W. designed the optical setup. K.J. implemented the data acquisition and DSP routines with the support of S.R. and C.R. \"O.B. provided the QRNG. K.J., T.D., B.H., and I.K. contributed to the temporal mode definitions. K.J. conducted the experiments, F.K. calculated the key rates and K.J. performed the final data analysis. S.P., T.G., B.\"O., M.H., and C.P. implemented the AIT-QPS pipeline. F.K., T.U., J.L., and N.L. contributed the security proof and provided theoretical support.
 G.L. and Ch.M. supervised the project. K.J., F.K., and M.H. wrote the manuscript with contributions from all authors.

\section*{Competing interests}
The authors declare no competing interests.

\section{Correspondence}
Correspondence and requests for materials should be addressed to
Kevin Jaksch or Christoph Marquardt.

\bibliographystyle{naturemag}
\bibliography{ref}

\newpage
\onecolumngrid
\section*{Supplementary note 1: Optimization problem for the security proof}
\onecolumngrid
In this section, we give additional information about the chosen observables, the acceptance set, and the semi-definite program we solve to obtain secure key rates. We start by introducing the observables used. While Alice's and Bob's heterodyne measurement allows them to measure the $x$ and $p$ quadrature of the incoming signals, for reasons of security analysis, it turns out that it is advantageous to use displaced versions of the photon number $\hat{n}$ and the squared photon number $\hat{n}^2$ as observables. We denote them by $\hat{n}_{\beta_i} := \hat{D}(\beta_i) \hat{n} \hat{D}^{\dagger}(\beta_i)$ and $\hat{n}_{\beta_i}^2 := \hat{D}(\beta_i) \hat{n}^2 \hat{D}^{\dagger}(\beta_i)$ for $i\in\{0,1,2,3\}$ and $\beta_i := \sqrt{\eta} \alpha_i$. $\hat{D}(\gamma)$ is the displacement operator. Note that this does not mean we need to know the systems' $\eta$ or $\alpha_i$ exactly and beforehand. We simply use the expectation according to our model and apply the displacement to obtain our displaced observables. In case the actual (unknown) values differ, we will see this in our observations. Those observables are directly related to the heterodyne measurement outcomes (see, for example, Ref. \cite{UpadhyayaThesis2021}).

Then, following the theory in Ref. \cite{Kanitschar_Thesis_2022, Kanitschar2023}, we arrive at the following optimization problem

\begin{align}
\begin{aligned}\label{eq:PrimalProblem}
    \beta :=& \min~ f(\bar{\rho})\\
    \text{s.t. }& \\
    & \mathrm{Tr}\left[P\right] + \mathrm{Tr}\left[N\right] \leq 2 \sqrt{w}\\
    & P \geq \mathrm{Tr}_{B}\left[ \overline{\rho} \right] - \rho_A\\
    & N \geq - \left( \mathrm{Tr}_{B}\left[ \bar{\rho} \right] - \rho_A \right) \\
    &\mathrm{Tr}\left[\left( \ket{j}\!\bra{j}\otimes\hat{n}_{\beta_j}\right) \bar{\rho} \right] \geq \mu_j + \langle \hat{n}_{\beta_j} \rangle - w ||\hat{n}_{\beta_j}||_{\infty} \\
    & \mathrm{Tr}\left[\left( \ket{j}\!\bra{j}\otimes\hat{n}_{\beta_j}\right) \bar{\rho} \right] \leq \mu_j + \langle \hat{n}_{\beta_j} \rangle  \\
    &\mathrm{Tr}\left[\left( \ket{j}\!\bra{j}\otimes\hat{n}_{\beta_j}\right) \bar{\rho}\right] \geq -\mu_j + \langle \hat{n}_{\beta_j} \rangle - w ||\hat{n}_{\beta_j}||_{\infty} \\
    & \mathrm{Tr}\left[\left( \ket{j}\!\bra{j}\otimes\hat{n}_{\beta_j}\right) \bar{\rho}\right] \leq -\mu_j + \langle \hat{n}_{\beta_j} \rangle  \\ 
    & \mathrm{Tr}\left[\left( \ket{j}\!\bra{j}\otimes \hat{n}^2_{\beta_j} \right) \bar{\rho}\right] \geq \mu_j + \langle \hat{n}_{\beta_j}^2 \rangle - w ||\hat{n}_{\beta_j}^2||_{\infty}\\
    &\mathrm{Tr}\left[\left( \ket{j}\!\bra{j}\otimes \hat{n}^2_{\beta_j} \right) \bar{\rho} \right] \leq \mu_j + \langle \hat{n}^2_{\beta_j} \rangle \\
    & \mathrm{Tr}\left[\left( \ket{j}\!\bra{j}\otimes\hat{n}^2_{\beta_j}\right) \bar{\rho} \right] \geq -\mu_j + \langle \hat{n}_{\beta_j}^2 \rangle - w ||\hat{n}_{\beta_j}^2||_{\infty} \\
    & \mathrm{Tr}\left[\left( \ket{j}\!\bra{j}\otimes\hat{n}^2_{\beta_j}\right) \bar{\rho} \right] \leq -\mu_j + \langle \hat{n}^2_{\beta_j} \rangle  \\ 
    & 1-w \leq \mathrm{Tr}\left[\overline{\rho}\right] \leq 1\\
    & \bar{\rho}, P, N \geq 0
\end{aligned}
\end{align}

where $j \in \{0,..., N_{\mathrm{St}}-1\}$. From the acceptance testing theorem (Theorem 4 in \cite{Kanitschar2023}), we obtain $\mu_X := \sqrt{\frac{||X||_{\infty}^2}{2m_{X}} \ln\left( \frac{2}{\epsilon_{\mathrm{AT}}} \right)}$ for $X \in \{\hat{n}_{\beta_i}, \hat{n}_{\beta_i}^2\}$ and we obtain as a function of the bounded detection range $M$ (see also Ref. \cite{Kanitschar2023}) the following $||\hat{n}_{\beta_i}||_{\infty} = M^2-\frac{1}{2}$, and $||\hat{n}_{\beta_i}^2||_{\infty} = M^4-\frac{1}{2}M^2$.

\newpage
\onecolumngrid
\section*{Supplementary note 2: Measured observables}

\onecolumngrid
\begin{table*}[h]

\begin{tabular}{*9c}
\toprule
Run &             $n\:[10^8]$ &        $N_\mathrm{tot} \:[10^9]$&         $T\:[1]$ &   $\eta\:[1]$ &       $\nu_{\mathrm{el}}\:[\mathrm{SNU}]$ &        $I_T\:[10^{-8}]$ &      $\mathrm{BER}\,\,X\:[1]$ &      $ \mathrm{BER}\,\,P\:[1]$ \\
\midrule
1 &  8.9866 &  1.1982 &  0.4950 &  0.720 &  0.1350 &  0.6677 &  0.3378 &  0.3368 \\
2 &  8.9866 &  1.1982 &  0.4950 &  0.720 &  0.1351 &  0.0000 &  0.3376 &  0.3368 \\
3 &  8.9866 &  1.1982 &  0.4959 &  0.720 &  0.1354 &  0.6677 &  0.3367 &  0.3357 \\
4 &  8.9866 &  1.1983 &  0.4938 &  0.720 &  0.1348 &  0.0000 &  0.3367 &  0.3362 \\
5 &  8.9866 &  1.1982 &  0.4943 &  0.720 &  0.1349 &  0.0000 &  0.3253 &  0.3245 \\
6 &  8.9866 &  1.1982 &  0.4914 &  0.720 &  0.1354 &  0.0000 &  0.3247 &  0.3261 \\
\bottomrule
\end{tabular}

\begin{tabular}{*5c}
\toprule
Run &              $\alpha_0\:[1]$ &               $\alpha_1\:[1]$ &               $\alpha_2\:[1]$ &               $\alpha_3\:[1]$ \\
\midrule
1 &  0.5289+0.5255j &  0.5338-0.5442j & -0.5343+0.5444j & -0.5286-0.5257j \\
2 &  0.5263+0.5221j &  0.5312-0.5414j & -0.5314+0.5420j & -0.5263-0.5215j \\
3 &  0.5289+0.5255j &  0.5338-0.5442j & -0.5343+0.5444j & -0.5286-0.5257j \\
4 &  0.5657+0.5678j &  0.5862-0.5745j & -0.5860+0.5745j & -0.5660-0.5681j \\
5 &  0.5657+0.5678j &  0.5862-0.5745j & -0.5860+0.5745j & -0.5660-0.5681j \\
6 &  0.5657+0.5678j &  0.5862-0.5745j & -0.5860+0.5745j & -0.5660-0.5681j \\
\bottomrule
\end{tabular}

\begin{tabular}{*9c}
\toprule
Symbol / &  \multicolumn{4}{c}{$\langle \hat{n}_{\beta_i} \rangle\:[10^{-3}]$} & \multicolumn{4}{c}{$\langle \hat{n}^2_{\beta_i} \rangle \:[10^{-3}]$}\\
Run &     0 & 1 & 2 & 3 & 0 & 1 & 3 & 3 \\
\toprule
1 &  0.9458 &  0.4128 &  0.0691 &  1.1511 &   5.2876 &   6.9297 &   6.3640 &   6.8303 \\
2 &  0.9354 &  0.1293 &  0.0390 &  0.9096 &   6.1404 &   7.0781 &   7.1981 &   5.9385 \\
3 &  0.9458 &  0.4128 &  0.0691 &  1.1511 &   5.2876 &   6.9297 &   6.3640 &   6.8303 \\
4 &  1.0567 &  0.2228 &  0.2246 &  1.4379 &   7.4074 &   7.3594 &   7.2678 &   7.2334 \\
5 &  1.0567 &  0.2228 &  0.2246 &  1.4379 &   7.4074 &   7.3594 &   7.2678 &   7.2334 \\
6 &  1.0567 &  0.2228 &  0.2246 &  1.4379 &   7.4074 &   7.3594 &   7.2678 &   7.2334 \\
\bottomrule
\end{tabular}
\caption{Measured parameters for all six runs and four QPSK states as used for the numerical SDP problem in \cite{Kanitschar2023}. $n$: Total number of private states. $N_\mathrm{tot}$: Total number of sent states. $T$: Channel transmission. $\eta$: Receiver efficiency. $\nu_{\mathrm{el}}$: Detector noise. $I_T$: Relative outliers for energy test. $\mathrm{BER}\,\,X/P$: Measured BER after applying key map. $\alpha_i$: Sending amplitude for each QPSK state. $\langle \hat{n}_{\beta_i} \rangle$: Mean displaced photon number for each QPSK state. $\langle \hat{n}^2_{\beta_i} \rangle$: Mean displaced squared photon number for each QPSK state.}
\end{table*}

\end{document}